\documentclass{article}

\usepackage{graphicx}
\usepackage{graphicx}
\usepackage{amsmath}
\usepackage{amssymb}
\usepackage{tabularx}
\usepackage{xcolor}
\begin{document}

\title{Efficient method for localized functions using domain transformation and Fourier sine series}

\author{Raka Jovanovic$^{a,b}$, Sabre Kais$^{a,c}$, and Fahhad H. Alharbi$^{a,d}$ $^{\ast}$\thanks{$^\ast$ Email: falharbi@qf.org.qa \vspace{6pt}} \\
\vspace{6pt} $^{a}${\em{Qatar Energy and Environment Research Institute, Doha, qatar}} \\
$^{b}${\em{Institute of Physics, University of Belgrade, Pregrevica 118, Zemun, Serbia}} \\
$^{c}${\em{Department of Chemistry, Purdue University, West Lafayette, IN 47907 US}} \\
$^{d}${\em{King Abdulaziz City for Science and Technology, Riyadh, Saudi Arabia}} }

\maketitle

\begin{abstract}
An efficient approach to handle localized states by using spectral methods (SM) in one and three dimensions  is presented. The method consists of transformation of the infinite domain to the bounded domain in $(0, \pi)$ and using the Fourier sine series as a set of basis functions for the SM. It is shown that with an appropriate choice of transformation functions, this method manages to preserve the good properties of original SMs; more precisely, superb computational efficiency when high level of accuracy is necessary. This is made possible by analytically exploiting the properties of the transformation function and the Fourier sine series. An especially important property of this approach is the possibility of calculating the Hartree energy very efficiently. This is done by exploiting the positive properties of the sine series as a basis set and conducting an extinctive part of the calculations analytically. We illustrate the efficiency of this method and implement it to solve the Poisson's and Helmholtz equations in both one and three dimensions. The efficiency of the method is verified through a comparison to recently published results for both one and three dimensional problems.

\end{abstract}

\section{Introduction}

In atomistic calculations of molecules and clusters, it is essential to use localized basis sets, where the wavefunctions vanish exponentially at large distances. There are many used sets and the most common ones are Slater-type and Gaussian-type orbitals (STO and GTO respectively) \cite{QC01,QC02,QC03,QC04,QC05,QC06}. Practically, the sets dectate the accuracy and maximum possible size of the system to be studied. To improve the accuracy and the scalability of the calculations, there has been a growing interest in developing new sets \cite{AC01,AC02,AC03,AC04} and many of these are based on the robust real-space numerical methods. This is known as numerical atomic orbitals (NAO). Numerically, many real space methods were used for NAO such as finite difference (FD), finite element (FE), spectral methods, etc \cite{AC05,AC06,AC07}. In  practical applications and especially in the case of FD and FE, the solution in infinite domain is -in most cases- found by solving the equation on some finite domain where the computational window is truncated to be manageable. 

The problem with using FD and FE in the case that the computational window is large, is that a high number of elements is needed. As a consequence, the algebraic system to be solved gets very big. Because of this, the processing time and the required memory become very large and in many cases, the problem can not be calculated efficiently without using computers with great power. This is especially noticeable if we need a high level of accuracy or a higher number of dimensions of the problem. One more disadvantage of this approach is that in the case of slowly-decaying functions, it gives results of lower quality.

In cases when high accuracy is needed but we desire to solve a relatively small algebraic system, spectral methods (SMs) have been proven to be a good choice \cite{canuto2006,le2010,hesthaven2011,shen2011}. In this method, the goal is to represent the real-space solution of the differential equation as a sum of certain ``basis functions". In a sense, this is basically very similar to the conventional atomistic calculations. The commonly used basis functions are usually some form of Fourier series, Chebyshev and Hermite polynomials \cite{ShenReview,ChebyInfi,ShenCheby}, Jacobi functions \cite{Jacobi}, Wiener rational basis functions \cite{WeinerInfi}, and many other orthogonal basis sets \cite{ChebyBook,Novak,ShenReview}. The problem is finding the coefficients in the sum in order to satisfy the differential equation as well as possible. This can be done using several different approaches that are suitable for different problem specifics. In the  Galerkin method, the basis functions have the desired behavior at the boundaries; in the case of the Tau method, this is not a requirement.

SMs are generally easier to implement for bounded domains. This is due to the properties of the used basis function set. Previously, this type of methods has been also used on infinite and semi-infinite domains  \cite{ShenReview,alharbi2009a,alharbi2010a,Matsushima:1997:SMU:272736.272740}. This has been achieved by adopting various numerical techniques such as using suitable basis sets, truncating the numerical window, and forcing size scaling. Beside the aformentioned basis sets, an interesting approach has been presented by Guo and Xu, using a  mixed Laguerre-Legendre pseudo spectral scheme  for the problem of incompressible fluid flow in an infinite strip \cite{Ben-yu2004}. In practical application, a disadvantages of these approaches is that it is necessary to calculate some integrals over the infinite domain when calculating the coefficients. Many integrals can be calculated analytically and hence improve the efficiency. But, in many cases, this is not possible and the integrals are calculated numerically. As known, this is very expensive computationally.

In this work, we focus on the use of coordinate transformation or mapping to some finite domain and then use a suitable basis function set like Chebyshev polynomials or Fourier series. In practice, the mapping function and the original basis set create a new set of basis functions.  Mapping functions have been divided into three general groups by the way they behave at infinity; namely, logarithmic, algebraic and exponential mapping. A thorough analysis of these approaches can be seen in Boyd's book \cite{ChebyBook}. Algebraic and logarithmic domain mappings have been widely analyzed in combination with Chebyshev  polynomials \cite{Cain1984272,Grosch1977273}. Weideman-Cloot introduce the domain mapping using the $sinh$ as the transformation function \cite{Weideman:1990:SMM:100905.100957}.

A serious challenge with using domain mapping is that the new transformed partial differential equation (PDE) often become complex and cumbersome to deal with. Because of this fact, research has also been conducted to use non-classical orthogonal systems on infinite domains \cite{Christov} or to use some type of mapped orthogonal systems \cite{Guo,Wang2002374}. The stability and efficiency of SMs in unbounded domains using Laguerre functions \cite{ShenSiam2000} and  mapped Legendre functions \cite{Shen04erroranalysis} have been analyzed. An overview and comparison of using mapped Jacobi, Laguerre and Hermite functions is presented by Shen and Wang \cite{ShenReview}. In their article, they emphasize on the advantage of transforming basis functions instead of the domain, which is to a large extent related to the level of complexity of the transformed PDE. On the other hand, it could be said that transforming the initial PDE and using standard basis sets is a more ``natural method". We say this in the sense that familiar and widely used sets of basis functions can be applied as a "black box".

In our work, we focus on transforming the original PDE from an infinite domain to a bounded one; while taking care that the new equation does not become overly complicated. We also seek that after the transformation, it is easy to calculate the coefficients for the expansion using an appropriate set of basis functions. More precisely, we design the method so that the resulted system matrices are highly sparse. This is done by using a trigonometric transformation of the domain in combination with a Fourier sine series. This combination of transformation function and basis set frequently has a consequence that  the integrals can be analytically calculated due to the positive properties of the sine function. The small size of the algebraic system and its sparsity result in improving the efficiency as both the memory and computational time will be reduced. This is due to the fact that the needed memory and the computational time are proportional to the system size.

In the presented work, we implemented SMs combined with domain transformation. The $(-\infty, \infty )$ domain is transformed to the bounded domain $(0, \pi)$. This is an extension of the work of Matsushima and Marcus \cite{Matsushima:1997:SMU:272736.272740} and Cloot and Wiedeman \cite{Cloot1992398} to 3D, more precisely we use the same transformation function.  Since the problem of interest has vanishing boundary conditions at infinity, the basis functions should be 0 at the boundaries. So, it is possible to use the Fourier sine series as the basis functions, and calculate the coefficients using the Galerkin method in the new domain. This is a modification of the aforementioned work \cite{Cloot1992398, Matsushima:1997:SMU:272736.272740} where the full Fourier series is used to the specifics of the problem of our interest.

The goal of our work is to use this method in 3D, especially to solve Poisson's and Helmholtz equations, from which the electrostatic interaction energy (Hartree energy) and its screened version can be calculated. We show that for the PDEs of interest, the proposed method is very efficient and consistent. This is done by exploiting the properties of the sine series which makes it possible to solve the underlying integrals analytically. We compare our method to recently published results using spectral methods based on Hermite  and mapped Laguerre functions  for different decaying behaviors at infinity for one dimensional (1D) case. Furthermore, we show show how to extend the method to 3D and conduct a comparison to recent results published for 3D problems.

The article is organized as follows. In the second section, the formulations are presented in multiple subsections. The method is implemented and the obtained results are presented and discussed in the third section. Many comparisons are conductued with various published works in 1D and 3D. In the case of 1D, various decaying behaviors at infinity are assumed and analyzed. Finally, we close with a concluding remarks.

\section{Formulation}

In this work, we propose solving physical PDEs in infinite 1D or 3D domains using domain transformation and Fourier sine series. In general, many of  the infinitely extended physical quantities vanish at infinities. The proposed solution is applicable to a problem with general form presented in 1D case
\begin{equation}
	\label{GenForm}
	\mathcal{\hat{L}} \; u(\textbf{x}) = f(\textbf{x})
\end{equation}
where $\mathcal{\hat{L}}$ can be any ordinary differential operator applied on the unknown function $u(\textbf{x})$ to result in the force $f(\textbf{x})$. The assumptions are:
\begin{itemize}
\item Both $u(\textbf{x})$ and $f(\textbf{x})$ vanish smoothly at infinities.
\item $f(\textbf{x})$ is $\mathbb{C}^k$ continuous where $k$ is the highest differential order in $\mathcal{\hat{L}}$.
\item $\textbf{x}$ is in the physical space.
\end{itemize}

In this paper, we will limit our selves to differential operators of order 2. The presented approach is though not limited and can be applied to higher order differential operator. Then, Eq. \ref{GenForm} takes the following form given in 1D case.

\begin{equation}
    \label{BasicForm}
    L_2(x) \; \frac{d^2u}{dx^2} + L_1(x) \; \frac{du}{dx}+L_0(x) \; u(x)= f(x)
\end{equation}

\subsection{Domain transformation}

We can define a transformation function
\begin{equation}
	\label{Transformation}
	x_i = g_i(y_i)
\end{equation}
where $x_i$'s are the components of the physical space point $\textbf{x}$ and $y_i$'s are the associated components in the computational space point $\textbf{y}$.
There are many possibilities to transform $x_i \in(-\infty,\infty)$ into $y_i \in(0,\pi)$. In the same way when working with a 3D problem, the transformation functions transfer the whole physical space into a cube between $\textbf{y}=(0,0,0)$ and $\textbf{y}=(\pi,\pi,\pi)$. In this subsection and some of the following ones, we will focus on the 1D case since it can be directly extended to more dimensions. So, the subscript ``$i$" will be ignored.

By domain transformation, the  derivatives are consequently transformed to

\begin{equation}
	\frac{d}{dx} = \frac{1}{\frac{dx}{dy}}\frac{d}{dy}
\end{equation}
Here listed are how, the lowest 3 differential orders will be transformed.

\begin{equation}
	f(x) \equiv f(y)
\end{equation}
\begin{equation}
	\frac{df}{dx} = \frac{1}{\frac{dx}{dy}}\frac{df}{dy}
\end{equation}
\begin{equation}
	\frac{d^2 f}{dx^2} = \frac{1}{\left(\frac{dx}{dy}\right)^2} \frac{d^2 f}{dy^2} + \frac{1}{\frac{dx}{dy}} \frac{d}{dy} \left( \frac{1}{\frac{dx}{dy}} \right) \frac{df}{dy}
\end{equation}

Similarly, all the $L$'s functions (in Eq. \ref{BasicForm}) are transformed. So,
\begin{equation}
	L_i(x) \equiv L_i(y)
\end{equation}
By applying the transformation and the new form of the derivatives, Eq. \ref{BasicForm} (in 1D) becomes
\begin{equation}
\label{BasicTrasformed}
  L_2(y) \; \frac{1}{\left(\frac{dx}{dy}\right)^2} \frac{d^2 u}{dy^2} + \left[ L_1(y) \; \frac{1}{\frac{dx}{dy}} + L_2(y) \; \frac{1}{\frac{dx}{dy}} \frac{d}{dy} \left( \frac{1}{\frac{dx}{dy}} \right)\right] \frac{du}{dy}+L_0(y) \; u(y)= f(y)
\end{equation}

In this work, we use the following trigonometric transformation from physical to computational space:
\begin{equation}
\label{TransFun}
y = \frac{\pi}{2}+\arctan (x)
\end{equation}
It can also be written in the inverse form, as transformation from computational to physical space and the associated derivatives are simply
\begin{equation}
	x = \tan (y-\frac{\pi}{2}) \; ,
\end{equation}
\begin{equation}
	\frac{1}{\frac{dx}{dy}} = \sin^2(y) \; ,
\end{equation}
and
\begin{equation}
{\frac{d}{dy}}\left( \frac{1}{\frac{dx}{dy}} \right)=2\sin(y) \cos(y) = \sin(2y)
\end{equation}

Other transformation functions can also prove to be efficient. This transformation allows the use of Fourier sine series to solve the problem, where any transformed physical quantity can be expanded as follow
\begin{equation}
    \label{ExpansionTransformed}
	f(x) \equiv f(y) = \sum_{m} c_{m} \sin(m y)
\end{equation}
Using Eq. \ref{ExpansionTransformed}, we can easily get the real space expansion in the following form:
\begin{equation}
\label{ExpansionReTransformed}
	f(x) = \sum_{m} c_{m} \sin (m (\frac{\pi}{2}+\arctan (x))).
\end{equation}

\subsection{Numeric solution}

The problem will be solved based on the aforementioned domain transformation and Fourier sine series. The problem will be first transformed from real-space to computational-space. Then, the problem will be solved in the computational space and later re-transformed back to the real-space.

In the computational space, the unknown function $u(y)$ is expanded in 1D using Eq. \ref{ExpansionTransformed}. By applying this expansion form and using moments calculation, Eq. \ref{BasicTrasformed} can be rewritten as
\begin{equation}
\label{MatrixForm}
\textbf{M} \; \textbf{c} = (\textbf{M}_2+ \textbf{M}_{12}+ \textbf{M}_1+ \textbf{M}_0 )  \; \textbf{c} = \textbf{f}
\end{equation}
where the elements of the \textbf{M} matrices are
\begin{equation}
    \label{M2}
	[\textbf{M}_2]_{mn}  = -n^2 \int_0^\pi \sin(my) \; L_2(y) \frac{1}{\left(\frac{dx}{dy}\right)^2} \sin(ny)\; dy
\end{equation}

\begin{equation}
	 \label{M12}
	[\textbf{M}_{12}]_{mn}  = n \int_0^\pi \sin(my) \; L_2(y)\frac{1}{\frac{dx}{dy}}{\frac{d}{dy}}{\left(\frac{1}{\frac{dx}{dy}}\right)} \cos(ny) \; dy
\end{equation}

\begin{equation}
	 \label{M1}
	[\textbf{M}_{1}]_{mn}  = n \int_0^\pi \sin(my) \; L_1(y)\frac{1}{\frac{dx}{dy}} \cos(ny)\; dy
\end{equation}

\begin{equation}
	 \label{M0}
	[\textbf{M}_{0}]_{mn}  = \int_0^\pi \sin(my) \; L_0(y) \sin(ny)\; dy
\end{equation}
$\textbf{c}$ is the vector of unknown coefficients
\begin{equation}
\textbf{c} = (c_1, c_2, .., c_m)^\top
\end{equation}
and $\textbf{f}$ is a vector containing the projection moments of the force function $f(y)$ based on the expansion. So, $\textbf{f}$ is
\begin{equation}
\textbf{f} = (f_1, f_2, .., f_m)^\top
\end{equation}
for which the values $f_m$ are calculated as follow
\begin{equation}
    \label{fcoef}
	f_m  = \int_0^\pi \sin(my) \; f(y) \; dy
\end{equation}

As both $\textbf{M}$ and $\textbf{f}$ are known from the expansion form and the system inputs (i.e. $f(x)$ and $L_i(x)$'s), Eq. \ref{MatrixForm} can be solved to find \textbf{c}, which contains the expansion coefficients of unknown function $u(x)$.

For Poisson's equation, which takes the following general form:
\begin{equation}
	\label{PoissonForm}
	 \nabla^2 u = f
\end{equation}
$L_2=1$ and $L_1=L_0=0$, and consequently,
$\textbf{M}_1=\textbf{M}_0=0$. It is also of great importance that  both $\textbf{M}_2$ and $\textbf{M}_{12}$ can be found analytically using the proposed transformation and basis set and they have the following forms:

\begin{equation}
	\left[ \textbf{M}_{12} \right]_{mn} = \frac{n \pi}{16} \left[ -\delta_{m-n,4} + \delta_{n-m,4} -\delta_{m+n,4} +2\delta_{m-n,2} -2\delta_{n-m,2} +2\delta_{m+n,2} \right]
\end{equation}
and
\begin{equation}
	\left[ \textbf{M}_2 \right]_{mn} = \frac{-n^2 \pi}{32} \left[ \delta_{m-n,4} + \delta_{n-m,4} -\delta_{m+n,4} -4\delta_{m-n,2} -4\delta_{n-m,2} +4\delta_{m+n,2} +6\delta_{m,n} \right]
\end{equation}
where $\delta_{i,j}$ is the Kronecker delta.

\subsection{Extension to Three Dimensions}

In the case of the 3D Poisons equation, the domain $(-\infty, \infty, ) \times (-\infty, \infty)\times (-\infty, \infty)$  is transformed to $(0, \pi) \times(0, \pi) \times (0, \pi)$. As in the case of 1D, the Fourier sine series are used to approximate the transformed function. So, the unknown function $u(x_1, x_2, x_3)$ is expanded as (just extension from 1D formulation)
\begin{equation}
u(y_1,y_2)= \sum_{lmn} c_{lmn} \sin(l y_1) \sin(m y_2)\sin(n y_3)
\end{equation}
In 3D, the general problem form (Eq. \ref{GenForm}) is recalled and it is
\begin{equation}	
\label{Lin2D}
\mathcal{\hat{L}} \; u(x_1,x_2,x_3) = f(x_1,x_2,x_3)
\end{equation}							
where $\mathcal{\hat{L}}$ can be any ordinary differential operator applied on the unknown function $u(x_1,x_2,x_3 )$ to result in the force $f(x_1,x_2,x_3)$. They are all defined in $(-\infty, \infty)\times(-\infty, \infty)\times(-\infty, \infty)$ where $x_1,x_2,x_3$ are the real-space parameters. We can defined a transformation functions
\begin{equation}
(y_1,y_2,y_3)= (g(x_1),g(x_2),g(x_3))
\end{equation} 								
where $y_1,y_2,y_3$ are the computational window parameters and the window is the cube $(0,\pi)\times(0,\pi)\times(0,\pi)$. $g$ is the transformation function used in 1D. This transformation changes obviously the functional form of the operator $\hat{L}$. Again, we will limit the differential order to 2 as we did in 1D case. So, Eq. \ref{Lin2D} is reduced to
\begin{multline}
    \label{BasicForm2D}
    L_2(x_1,x_2,x_3) \left(\frac{d^2u}{dx_1^2}+\frac{d^2u}{dx_2^2} +\frac{d^2u}{dx_3^2}\right) + L_1(x_1,x_2,x_3) \left(\frac{du}{dx_1}+\frac{du}{dx_2}+\frac{du}{dx_3}\right)\\ +L_0(x_1,x_2,x_3) \; u(x_1,x_2,x_3)= f(x_1,x_2,y_3)
\end{multline}

By using the same number of bases to all coordinates, Poisson's equation can be rewritten in a matrix form as follow:

\begin{equation}
\label{Matrix2D}
\left( \textbf{I} \otimes\textbf{I} \otimes \textbf{M}_{1D}   + \textbf{I} \otimes \textbf{M}_{1D}\otimes \textbf{I}+ \textbf{M}_{1D} \otimes \textbf{I} \otimes \textbf{I} \right) \textbf{c} = \textbf{M}_{3D} \; \textbf{c} = \textbf{f}
\end{equation}
where $\otimes$ represents the Kronecker product, $\textbf{M}_{1D}$ is the corresponding matrix for the 1D Poisson's Equation, and $\textbf{I}$ is the identity matrix. This equation can be solved directly to find the coefficients $\textbf{c}$, which is the vector of unknown coefficients for the 3D problem and is defined as
\begin{equation}
\textbf{c} = (c_{111}, c_{211}, .., c_{nnn})^\top
\end{equation}
and $\textbf{f}$ is the force vector and it is defined as well as
\begin{equation}
\textbf{f} = (f_{111}, f_{211}, .., f_{nnn})^\top
\end{equation}
for which the values $f_{lmn}$ are calculated as in Equation \ref{fcoef2d}.
\begin{equation}
    \label{fcoef2d}
	f_{lmn}  = \int_0^\pi \int_0^\pi \int_0^\pi \sin(ly_1) \; \sin(my_2)\sin(ny_3) \; f(y_1,y_2,y_3) \; dy_1 dy_2 dy_3
\end{equation}

\subsection{Calculating Hartree energy}

Hartree energy $E_H$ is

\begin{equation}
\label{Hartree}
	E_H = \int_\Omega \int_\Omega \frac{\rho(\textbf{x}_1)\rho(\textbf{x}_2)}{|\textbf{x}_1-\textbf{x}_2|}  d^3 x_1 d^3 x_2 = \int_\Omega  V(\textbf{x})\rho(\textbf{x}) d^3 x
\end{equation}
where $\Omega$ is the whole physical space and $V(\textbf{x})$ is the solution of Poisson equation

\begin{equation}
	\label{PoissonForm2}
	 \nabla^2 V(\textbf{x})  = - 4\pi \rho(\textbf{x})
\end{equation}

In general, to calculate $E_H$, we need to calculate $N^3$ integrals, where $N$ is the number of the used basis set in each domain. This calculation is computationally very expensive. As we have previously mentioned using Eq. \ref{ExpansionReTransformed}, we can represent $V(\textbf{x})$ and $\rho(\textbf{x})$ as

 \begin{equation}
	V(\textbf{x}) = \sum_{lmn} c_{lmn} \sin(l (\frac{\pi}{2}+\arctan (x_1))) \sin(m (\frac{\pi}{2}+\arctan (x_2))) \sin(n (\frac{\pi}{2}+\arctan (x_3)))
\end{equation}

\begin{equation}
	\rho(\textbf{x}) = \dfrac{-1}{4 \pi} \sum_{lmn} f_{lmn} \sin(l (\frac{\pi}{2}+\arctan (x_1))) \sin(m (\frac{\pi}{2}+\arctan (x_2))) \sin(n (\frac{\pi}{2}+\arctan (x_3)))
\end{equation}
Using these expansions, Eq. \ref{Hartree} is reduced to
\begin{equation}
	E_H =  \textbf{c}^\top \textbf{E}_{3D} \textbf{f}
\end{equation}
and
\begin{equation}
	\textbf{E}_{3D} =  \textbf{E} \otimes \textbf{E} \otimes \textbf{E} \; .
\end{equation}
The elements of the matrix $\textbf{E}$ are
\begin{equation}
	[E]_{i,j} = \int_{0}^{\pi} \sin(m y)\sin(ny) \frac{1}{\sin(y)^2} dy = \pi \mod(|i-j|+1,2) \min(i,j).
\end{equation}
where ``$\mod$" and ``$\min$" are the modular arithmetic and the minimum respectively. It is clear that by exploiting the properties of the Fourier sine series we have a great speed up in calculating the Hartree energy since no new integrals need to be calculated.

\section{Implementation, Results, and Discussions}

In the first subsection, we give an analysis of the efficiency of applying the proposed method for problems with known solutions. We observe the speed of convergence towards the exact solution depending on the number of basis functions in the spectral method for different behavior at infinity. The efficiency of this method is evaluated by comparing it to recently published results to a slightly more complex PDE. The second group set of tests is dedicated to the 3D version of the problem where the presented method is compared to recently published results. Several practical problems have been encounter during the implementation of this method which will be also mentioned. The method has been implemented by creating code using MatLab R2013a and the calculations have been done on a machine with Intel(R) Core(TM) i7-2630 QM CPU \@ 2.00 GHz, 4GB of DDR3-1333 RAM, running on Microsoft Windows 7 Home Premium 64-bit.
\subsection{Comparison to other methods in 1D}
\label{OtherMethods}
To better evaluate the proposed method, we compare it to recently published results by Shen and Wang \cite{ShenReview}. Of course, we have only included problems in the domain $(-\infty, \infty)$. In our comparison, the following 1D equation is solved.

\begin{equation}
	\label{TestProblem}
	 -\nabla^2 u + \gamma u= f
\end{equation}
For the new equation, only minor changes need to be done compared to the Poisson's equation. More precisely, when calculating $\textbf{M}$ for Eq. \ref{MatrixForm}, we have to take into account the new term $\gamma u$. In this case, $L_2$ and $L_0$ are set to -1 and $\gamma=2$ respectively in Eq. \ref{BasicTrasformed}. Practically,
\begin{equation}
	\label{TestForm}
	 \textbf{M}_0= \gamma\frac{\pi}{2}\textbf{I}
\end{equation}

As proposed by Shen and Wang, we shall observe the accuracy of the method for different decay properties of function $u$. More precisely, we analyze exponential decay with oscillation at infinity with the following representative function
\begin{equation}
	\label{DecExpOsc1}
	u(x) =  \sin(kx)e^{-x^2},
\end{equation}
algebraic decay given by the following equation
\begin{equation}
	\label{DecAlg}
	u(x) =   \frac{1}{(1+x^2)^h},
\end{equation}
and algebraic decay with oscillation as follow
\begin{equation}
	\label{DecAlgOsc}
	u(x) =   \frac{\sin(kx)}{(1+x^2)^h}
\end{equation}

\begin{figure}
\centering
\includegraphics[width=.85\textwidth]{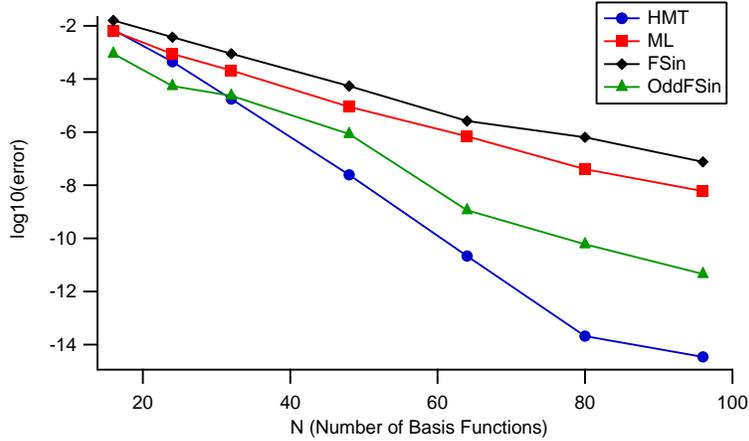}
\caption{ Convergence rates with exact solution  using error in maximum norm for $u$ with exponential decay and oscillation (Eq. \ref{DecExpOsc1}, k=2). }
\label{SinExp}
\end{figure}

\begin{figure}
\centering
\includegraphics[width=.85\textwidth]{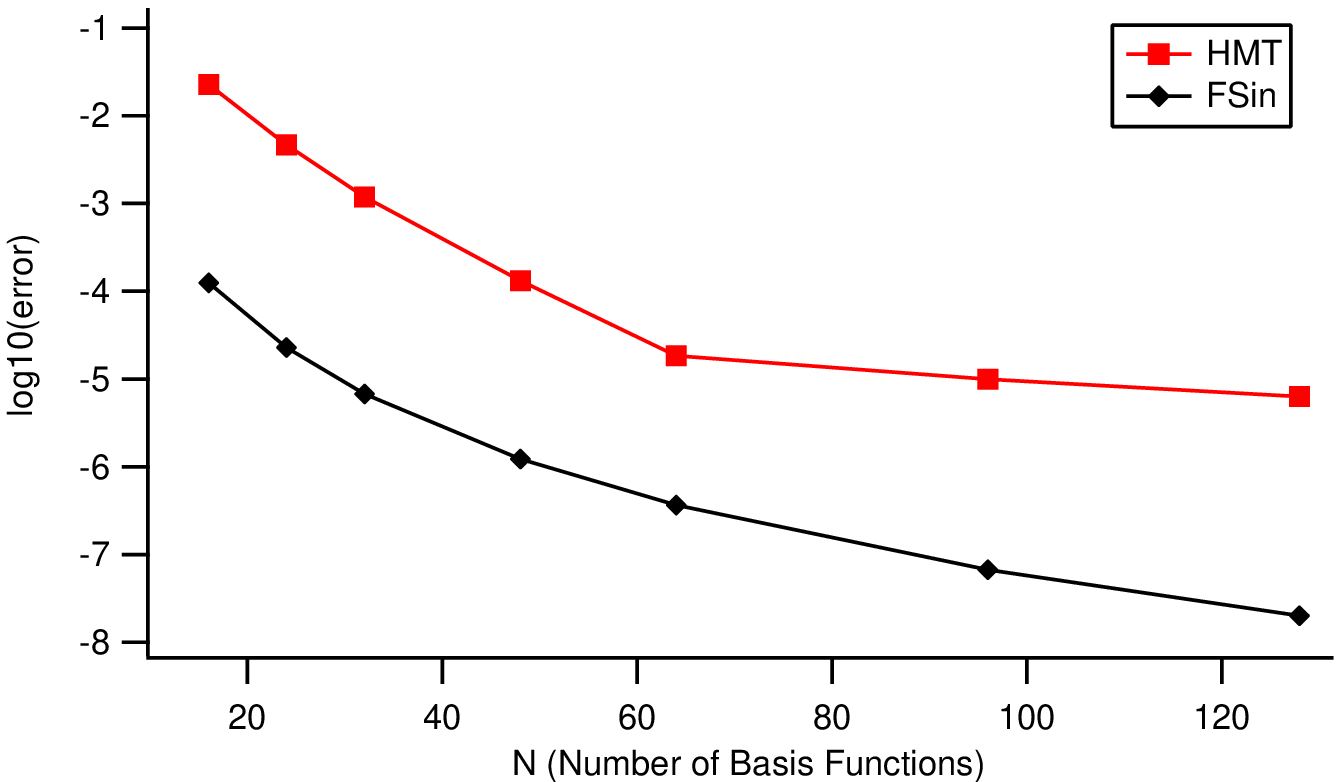}
\caption{ Convergence rates with exact solution  using error in maximum norm for $u$ with algebraic decay  (Eq. \ref{DecAlg},  h=2).}
\label{AL}
\end{figure}

\begin{figure}
\centering
\includegraphics[width=.85\textwidth]{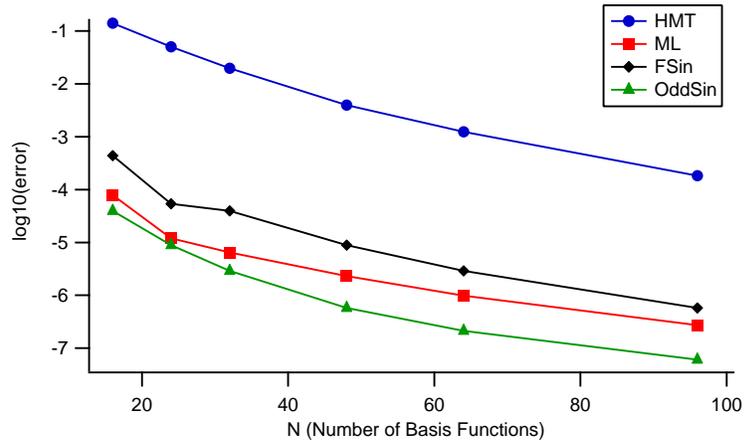}
\caption{ Convergence rates with exact solution  using error in maximum norm for $u$ with algebraic decay and oscillation (Eq. \ref{DecAlgOsc}, h=$ \frac{7}{2}$, k=2). }
\label{SinAL}
\end{figure}

In the tests,  $u(x)$ has been taken to be the exact solution of Eq. \ref{TestProblem}, and the corresponding value for $f(x)$ has been calculated analytically. We compare the accuracy of approximate solutions acquired by our method using Fourier sine series (FSin), to ones calculated using Hermite functions (HMT) or mapped Laguerre functions (ML). The values for HMT and ML have been taken from article \cite{ShenReview}. The results can be seen in Figures \ref{SinExp}, \ref{AL} and \ref{SinAL}.

In all of the examples, we have compared the error to the exact solution using error in maximum norm. We wish to point out, that in the cases when the exact solution $u$ was an odd functions as in Eqs. \ref{DecExpOsc1}, \ref{DecAlgOsc}, it is possible to exploit this fact. It is well known that the Fourier sine series for odd functions has all the even coefficients equal to zero. In practical applications, in many cases the parity of the function that we wish to approximate is known in advance, or it can be deducted from $f$ in Eq. \ref{TestProblem}. Because of this, we have also added  to our comparison the accuracy of our method when we use only odd members of the Fourier sine series (OddFSine).

By observing the results in Figures \ref{SinExp}, \ref{AL} and \ref{SinAL}, we can first observe that our method has a robust behavior compared to the other two methods. Using the sine series combined with a transformation always produces results of accuracy close to $10^{-6}$.  Contrary to this, in the case of algebraic decay the HMT  has a bad performance in both cases,  with or without oscillation. In Figure \ref{AL}, the results for ML are not included because they where not presented in \cite{ShenReview}. The function $u$ that has been used for comparison of the two methods in \cite{ShenReview} had the value $h=3.5$ in Eq. \ref{DecAlg}. This value of $h$, was an exceptional case for our method and we would get accuracy of $10^{-15}$ within 5 basis functions, because of this it was excluded from 
the analysis.

If the odd property of $u$ was exploited, our method would  get higher accuracy than ML in all the tested cases. We point this out since ML uses the concept of mappings which are mathematically equivalent to transformations used in our work. In case of ML, different mappings are used depending on the type of decay (exponential or algebraic); while in our case, the same transformation function gives good results in all the tests. This is a clear advantage since the decaying behavior at infinity is not always known. We believe that the transformation function given in Eq. \ref{TransFun} can be adapted to be more suitable to different decay properties but this is out of the scope of this article.
\subsection{Comparison to other methods in 3D}

In this section we compare our results to the non iterative method for solving Poisson's and Helmholtz equation given in  article \cite{Berger2005235}. In the article by Berger and Sundholm, the 3D version of Eq. \ref{TestProblem}, has been solved for $\gamma = -1, 0, 1$.  In the tests, the same  $N$ number of basis functions have been used in all three dimension, which means that $N^3$ basis functions have been used in total. The matrix $M_{3D}$ from Eq. \ref{Matrix2D}, as in the 1D case is very sparse. In our tests we have experienced that in 1D case, the greatest part of calculation time is used for calculating the coefficients $f_{i}$ given in Eq. \ref{fcoef}. In practical application for 3D problems, it is hard to efficiently calculate numerically the 3D integrals from Eq. \ref{fcoef2d}, due to very
 long execution time. In general, when using this method it is necessary to give special consideration to this calculation. This can -in many cases- be done by trying to separate the variables and calculate 1D integrals. This is simply done in the case of our test example. Another problematic area with practical application of this method is solving the algebraic equation for finding the expansion coefficients due to its gigantic size $(N^3)$.
   So although the method it self is not iterative, due to memory restrictions we had to use an iterative method for solving the underling algebraic system.

 Berger and Sundholm tested their method for an example with function $\rho(\textbf{x})$ having the following value:
\begin{equation}
        r^2 = \textbf{x} \cdot \textbf{x} \\
\end{equation}
 \begin{equation}
        \rho(x_1,x_2,x_3)= x_1 x_2 x_3(6+ 4r^2)e^{-r^2}
\end{equation}
For evaluating their  method they calculated $E_H$. We show the comparative results in Table \ref{table:3D}.  The comparison has been done between the same number of grid points in case of article \cite{Berger2005235}, and number of basis functions for our method.

\begin{table}[h]
\footnotesize
\center
\caption{\label{table:3D} $E_H$ as a function of grid and expansion sizes obtained by solving Poisson equation ($k^2=0$) and Helmholtz equation ($k^2=1$).}
\begin{tabular}{|c| c| c| c| c|}
\hline
Grid &  0 Berger \cite{Berger2005235} & FSin & 1 Berger \cite{Berger2005235} & FSin\\
\hline
15 x 15 x 15   &  - & 1.3169389525& -  &1.0799669634\\
25 x 25 x 25   &  1.4374409362 &1.3314783170 &1.1784597644&1.0867919307\\
37 x 37 x 37   &  1.3287968899 &1.3315096147&1.0842696689&1.0868280822\\
49 x 49 x 59   &  1.3315672651 &1.3315099559& 1.0868529977 &1.0868283787\\
61 x 61 x 61   &  1.3317110807 &1.3315099954& 1.0869883397 &1.0868284661\\
73 x 73 x 73   &  1.3314981186 &1.3315099945 & 1.0868078129 &1.0868284685\\
85 x 85 x 85   &  1.3315254242 &1.3315099968& 1.0868316094  &1.0868284686\\
97 x 97 x 97   &  1.3315234616 &1.3315099905& 1.0868293521  &1.0868284686\\
109 x 109 x 109&  1.3315223680 &- & 1.0868284706  &-\\
121 x 121 x 121&  1.3315223246 &- & 1.0868284564  &-\\
133 x 133 x 133&  1.3315223421 & -& 1.0868284711  &-\\
145 x 145 x 145&  1.3315223475 &- & 1.0868284726  &-\\

\hline
\end{tabular}
\end{table}

We can first notice that for both  cases of $\gamma=0, 1$ our method has a much faster convergence speed in respect of the size of the algebraic system that needs to be solved. This is an important advantage since a frequent limitation with solving PDE's is the lack of available memory.
 We would also like to point out the slight difference in results between the two methods. We believe that the difference comes from the fact that Berger and Sundholm only solve the PDE on the truncated domain (-10, 10),  and uses its value to calculate $E_H$ on the infinite domain. The negative effect of using a truncated domain on the precision of calculating the Hartree energy has been presented on a similar problem in article  \cite{lee}. Contrary to this with our approach, a great part of the calculation is done analytically, which gives us an arguments to say our method is more precise.

We have excluded from our results the case of $\gamma=-1$, because our method did not manage to converge to the solution. This behavior is not surprising, since for $\gamma=-1$, we have the reduced wave equation, which is significantly harder to solve than for $\gamma=1$ which represents stationary reaction-diffusion phenomena \cite{HardHelmholtz}. Discretizations of the  Helmholtz equation in this case, using spectral element methods, can often result in linear system of equations which possesses an indefinite coefficient matrix \cite{HardHelmholtz}. The standard approach to resolving this problem is the use of some preconditioning to the matrix, but we believe that a detailed analysis of this aspect is out of scope of the article.

\section{Conclusion}

In this paper, we have presented an efficient method that solves 1D and 3D PDEs in infinite domains with vanishing boundary conditions (localized functions). The method is based on transforming the infinite domain to the bounded domain $(0, \pi)$ using a trigonometric function. The transformed PDE in the new finite domain is then solved  using SMs. The Fourier sine series has been selected as the basis function set. We have shown that the specified transformation has two main advantages. First it is possible to use the Galerkin method for finding the coefficients of the function expansion. Secondly, we show that by using the proposed transformation function, the algebraic system that needs to be solved is very sparse and can be easily calculated.

We have also illustrated  the efficiency of this approach by solving the Poisson's and Helmholtz equation in both 1D and 3D and calcualting Hartree energy for a realistic 3D cases. The conducted tests have shown that the proposed method is capable of finding solutions of high accuracy with a relatively small number of basis functions. We have analyzed the convergence speed of the method and the distribution of the error over the domain of interest. Some general guidelines are given for effective implementation of the proposed method. Further, when our approach had been compared to similar methods for problems with different decay properties, it has shown a more robust behaviour. This work is part of a longer term goal of developing an efficient orbital-free density functional theory tool using meshfree spectral method for real space. It will be used to calculate opto-electronic properties of atomic, molecular, and extended systems.

\bibliographystyle{tMPH}
\bibliography{LocStates}

\end{document}